\def\be{\begin{equation}}
\def\ee{\end{equation}}
\def\ba{\begin{array}}
\def\ea{\end{array}}

\documentclass[prl,showpacs,twocolumn,amsmath]{revtex4}
\usepackage{amsfonts}
\usepackage{mathrsfs}
\usepackage{mathdots}
\usepackage{amssymb}
\usepackage{pifont}
\usepackage{epsfig,subfigure,dsfont,amsthm,amsbsy,mathrsfs,amscd}
\usepackage{color}

\usepackage{rotating}

\usepackage{graphicx}
\usepackage{dcolumn}
\usepackage{bm}
\usepackage{amsmath}
\usepackage{multirow}
\usepackage{array}
\usepackage{epstopdf}
\usepackage{diagbox}
\usepackage{dcolumn}

\def\qed{\leavevmode\unskip\penalty9999 \hbox{}\nobreak\hfill
     \quad\hbox{\leavevmode  \hbox to.77778em{%
               \hfil\vrule   \vbox to.675em%
               {\hrule width.6em\vfil\hrule}\vrule\hfil}}
     \par\vskip3pt}

\newtheorem{proposition}{Proposition}

\begin{document}

\title{Genuine four-partite Bell nonlocality in the curved spacetime}
\author{Guang-Wei Mi$^{1}$}
\author{Xiaofen Huang $^{1}$}
\author{Shao-Ming Fei $^{2}$}
\author{Tinggui Zhang$^{1,*}$}
\affiliation{$^{1}$School of Mathematics and Statistics, Hainan Normal University, Haikou, 571158, China\\
$^{2}$School of Mathematical Sciences, Capital Normal University, Beijing, 100048, China\\
{$^{*}$ Email address: tinggui333@163.com}}

\date{\today}

\begin{abstract}
The quantum Bell nonlocality plays a crucial role in quantum information processing. In this paper, we first obtain the quantitative analytical expression of the genuine four-partite Bell nonlocality of any 4-qubit quantum state. Then, we investigate and derive the genuine four-partite nonlocalities in the background of the Schwarzschild black hole and the Schwarzschild-de Sitter spacetime. In the Schwarzschild black hole, it is shown that the Hawking effect destroys the accessible genuine four-partite nonlocality while simultaneously having both detrimental and beneficial impacts on the inaccessible genuine four-partite nonlocality. Besides, we observe that the Hawking effect can generate the physically inaccessible genuine four-partite nonlocality for fermion fields in Schwarzschild black hole. This result shows that the genuine four-partite nonlocality can pass through the event horizon of black hole. Moreover, we also investigate the influence of the Hawking effect of the Schwarzschild-de Sitter spacetime on the genuine four-partite nonlocality of massless Dirac fields.
\end{abstract}

\maketitle


\section{Introduction}
In 1964, Bell \cite{Bell1964} demonstrated that any physical theory that adheres to the principles of local realism is in conflict with the predictions of quantum theory, that is, the quantum Bell nonlocality. Quantum Bell nonlocality is an essential aspect of quantum mechanics and plays important roles in quantum information processing such as quantum key distribution~\cite{Barrett1991,Masanes2011}, communication complexity~\cite{Brukner2002,Buhrman2010} and quantum cryptography~\cite{Ekert1991,Bennett1992,AcAn2007}.

The quantum Bell nonlocality has been extensively investigated in bipartite scenarios. However, the researches in the multipartite scenarios are relatively limited ~\cite{ndsv,Ghose.2009,Ajoy.2010,Bancal.2011,Bancal.2013}. The resource of genuine multipartite nonlocality is of significance and has attracted many attention. In the tripartite scenario, the genuine tripartite nonlocality can be identified through the violation of the Svetlichny inequalities~\cite{Svetlichny1987}. In \cite{Seevinck2002} Michael Seevinck generalized the Svetlichny inequalities to the case of four or more partite systems. Moreover, Su \emph{et al.} \cite{Su2018} conducted a quantitative analysis of the genuine tripartite nonlocality. Then Wang \emph{et al.} \cite{Wang2020} derived an analytical expression of the genuine tripartite nonlocality. The analytical expression of genuine four or more partite nonlocality is less known.

In 1915, Einstein proposed the general theory of relativity, which predicted the existence of a celestial object called black hole. Subsequently, with the development of astronomy, the existence of black holes has been indirectly confirmed~\cite{Abbott2016,L12019,L172022}. However, there are still many unknown mysteries surrounding black holes. In 1974, Hawking discovered that black holes emit thermal radiation, which is known as Hawking radiation~\cite{Hawking1974}. This phenomenon connects quantum mechanics and gravity, and gives rise to the black hole information paradox~\cite{Hawking1975,Hawking1976,Bombelli1986}. Indeed, the black hole is asymptotically de Sitter, not asymptotically flat. A static and uncharged black hole is connected to the Schwarzschild-de Sitter (SdS) spacetime which describes the gravitational effects caused by both mass $M$ and the cosmological constant $\Lambda$. The SdS spacetime is obtained by combining solutions of the Schwarzschild black hole and the de Sitter universe. It describes the presence of a massive black hole in an expanding universe with cosmological constants. This spacetime has both a black hole event horizon (BEH) and a cosmological event horizon (CEH), which distinguishes it from single horizon spacetimes~\cite{Kottler1918,Stuch,Akcay2011}.

In recent years, aiming to better understand the unification of quantum mechanics and relativity, the study of quantum information in non-inertial frames and curved spacetime has become a new research hotspot~\cite{Schuller2005,Alsing2006,Wang2009,Wang2010,Wang2011,Esfahani2011,Tian2013,Xu2014,Kanno2016,Friis2016,Liu2018,Qiang2018,Class2020,Int2020,New2022,Eur2022,Li2022}.
In~\cite{Phy2024}, Wu \emph{et al.} explore the characteristics of genuine N-partite entanglement of Dirac fields and its restrictive relationships in the background of the Schwarzschild black hole. In~\cite{Wu2024}, Wu \emph{et al.} examine the properties of genuine N-partite entanglement of Dirac fields in SdS spacetime with the inclusion of the BEH and the CEH. Apart from the entanglement, the study on the nonlocality in the background of the Schwarzschild black hole and SdS spacetime is also of importance.
In~\cite{Eur2022,Li2022,Zhang.2023}, the authors study the the genuine tripartite nonlocality of Dirac fields in the Schwarzschild black hole.

In the paper, we first conduct a quantitative analysis and present an analytical expression of the genuine four-partite nonlocality. Then, we study the genuinely accessible and inaccessible four-partite nonlocality in Schwarzschild black hole. We initially assume that four observes share a four-partite GHZ-state in the asymptotically flat region. Next, we let $n$ $(n<4)$ observers hover near the event horizon of the black hole, while $4-n$ observers remain at the asymptotically flat region. Thus, we get the analytic expression that encompasses both physically accessible and inaccessible genuine four-partite nonlocality.
We obtain that the Hawking effect of the black hole destroys the accessible genuine four-partite nonlocality, while have both negative and positive impacts on inaccessible genuine four-partite nonlocality.
Moreover, we find that the Hawking effect can generate the physically inaccessible genuine four-partite nonlocality in the Schwarzschild black hole, that is, the genuine four-partite nonlocality can pass through the event horizon of black hole.
Finally, we consider the genuine four-partite nonlocality in the SdS spacetime with the BEH and CEH. Let $n$ observers near BEH and $m$ $(n+m=4)$ observers near CEH. We obtain an analytic expression of genuine four-partite nonlocality in the SdS spacetime. We conclude that for different modes, the $S(\rho_{4})$ has different properties with respect to the mass $M$ and the cosmological constant $\Lambda$, that is, the Hawking effect may either disrupt or enhance the genuine four-partite nonlocality.

The rest of this paper is organized as follows. In Sec. \uppercase\expandafter{\romannumeral2}, we give quantitative analysis on genuine nonlocality of general four-qubit states.
In Sec.\uppercase\expandafter{\romannumeral3}, we consider the genuinely accessible and inaccessible four-partite nonlocality in Schwarzschild black hole.
In Sec.\uppercase\expandafter{\romannumeral4}, we study the genuine four-partite nonlocality in the Schwarzschild-de Sitter black hole spacetime.
We conclude in Sec.\uppercase\expandafter{\romannumeral6}.

\section{GENUINE NONLOCALITY OF GENERAL FOUR-QUBIT STATE}

We first briefly review the four-partite Bell scenario and the Svetlichny inequality. Four parties, Alice(A), Bob(B), Charlie(C) and Dick(D), share a four-qubit quantum state. Each party selects one of two possible measurement outcomes $\pm1$ to perform a dichotomic projective measurement. Without loss of generality, assume that the two measurement observables for Alice are $A=\mathbf{a}\cdot\bm{\sigma}$ and $A^{\prime}=\mathbf{a^{\prime}}\cdot\bm{\sigma}$. Similarly, for Bob, Charlie and Dick, we have $B=\mathbf{b}\cdot\bm{\sigma}$, $B^{\prime}=\mathbf{b^{\prime}}\cdot\bm{\sigma}$, $C=\mathbf{c}\cdot\bm{\sigma}$, $C^{\prime}=\mathbf{c^{\prime}}\cdot\bm{\sigma}$ and $D=\mathbf{d}\cdot\bm{\sigma}$, $D^{\prime}=\mathbf{d^{\prime}}\cdot\bm{\sigma}$, respectively. Here, $\mathbf{a},\mathbf{a^{\prime}},\mathbf{b},\mathbf{b^{\prime}},
\mathbf{c},\mathbf{c^{\prime}},\mathbf{d},\mathbf{d^{\prime}}$ are all three-dimensional real unit vectors, and $\bm{\sigma}=\{\sigma_{1},\sigma_{2},\sigma_{3}\}$ is the vector given by the standard Pauli matrices. The four partite Svetlichny operator is given by ~\cite{Seevinck2002},
\begin{eqnarray}
\begin{aligned}
S&=[A\otimes B-A^{\prime}\otimes B^{\prime}]\otimes[(C-C^{\prime})\otimes D-(C+C^{\prime})\otimes D^{\prime}]\\
&-[A^{\prime}\otimes B+A\otimes B^{\prime}]\otimes[(C+C^{\prime})\otimes D-(C-C^{\prime})\otimes D^{\prime}],
\end{aligned}
\end{eqnarray}
and the Svetlichny inequality of a four-qubit state $\rho$ is given by
\begin{eqnarray}
tr(S\rho)\leq 8.
\end{eqnarray}

Any four-qubit state that violates this Svetlichny inequality is genuinely nonlocal. It has been shown that the maximum violation value of the Svetlichny inequality is $S_{max}=8\sqrt{2}$ \cite{Seevinck2002}, which is attained by the GHZ state $|GHZ\rangle=(|0000\rangle+|1111\rangle)/\sqrt{2}$. We define $S(\rho)$ to be the maximum violation of the Svetlichny inequality for a given state $\rho$,
\begin{eqnarray}
\begin{aligned}
S(\rho)\equiv \mathop{\max}\limits_{S}tr(S\rho),
\end{aligned}
\end{eqnarray}
where the maximum is taken over all possible Svetlichny operators. Let
\begin{eqnarray}
\begin{aligned}
N(\rho)=max\left\{0,\frac{S(\rho)-8}{S_{max}-8}\right\} \label{eqN}.
\end{aligned}
\end{eqnarray}
$N(\rho)$ monotonically increases from $0$ to $1$ as $S(\rho)$ ranges from $8$ to $S_{max}$. $N(\rho)=0$ if and only if $S(\rho)\leq8$, while $N(\rho)=1$ if and only if $S(\rho)=S_{max}$. That is, $N(\rho)$ is a comprehensive measure for quantifying the genuine nonlocality of the four-partite state $\rho$. If $N(\rho)>0$, the four-partite state $\rho$ is genuinely nonlocal. That is, When $8<S(\rho)\leq8\sqrt{2}$, $\rho$ is genuinely nonlocal.
To estimate $N(\rho)$, we calculate $S(\rho)$ in the following.

Any four-qubit state $\rho$ can be expressed in the Pauli matrix basis,
\begin{eqnarray}
\begin{aligned}
\rho=\frac{1}{16}\sum^{3}_{i,j,k,l=0}t_{ijkl}\sigma_{i}\otimes\sigma_{j}
\otimes\sigma_{k}\otimes\sigma_{l},\label{eq4-qubit}
\end{aligned}
\end{eqnarray}
where $t_{ijkl}=tr[\rho(\sigma_{i}\otimes\sigma_{j}\otimes\sigma_{k}\otimes\sigma_{l})]$, $\sigma_{0}$ denotes the $2\times 2$ identity matrix. We have
\begin{eqnarray}
\begin{aligned}
tr[\rho(A\otimes B\otimes C\otimes D)]&=\sum^{3}_{i,j,k,l=1}a_{i}b_{j}c_{k}d_{l}t_{ijkl}\\
&=\langle\mathbf{c},T_{ab}\mathbf{d}\rangle,
\end{aligned}
\end{eqnarray}
where $T_{ij}=(t_{ijkl})$ are $3\times 3$ real matrices with row and column indices given by $k$ and $l$, respectively, $i,j=1,2,3$, $T_{ab}=\sum^{3}_{i,j=1}a_{i}b_{j}T_{ij}$, and $\langle\cdot,\cdot\rangle$ represents the standard inner product in $\mathbb{R}^{3}$.

It is observed that the genuine nonlocality of a four-qubit state is solely determined by the $81$ real variables in matrices $T_{ij}$, $i,j=1,2,3$. The $9$ matrices $T_{ij}$, $i,j=1,2,3$, are called correlation matrices of the four-qubit state $\rho$.
Let $\mathbf{e_{0}}$ and $\mathbf{e_{1}}$ be two orthogonal unit vectors such that $\mathbf{c}+\mathbf{c^{\prime}}=2\cos\alpha\mathbf{e_{0}}$ and $\mathbf{c}-\mathbf{c^{\prime}}=2\sin\alpha\mathbf{e_{1}}$ for some $\alpha$. Let $E=\mathbf{e_{0}}\cdot\bm{\sigma}$ and $E^{\prime}=\mathbf{e_{1}}\cdot\bm{\sigma}$.
We have
\begin{eqnarray}
\begin{aligned}
tr(S\rho)&=2\cos\alpha tr\{\rho[(A^{\prime}\otimes B^{\prime}-A\otimes B)\otimes E\otimes D^{\prime}\\
&-(A^{\prime}\otimes B+A\otimes B^{\prime})\otimes E\otimes D]\}\\
&+2\sin\alpha tr\{\rho[(A\otimes B-A^{\prime}\otimes B^{\prime})\otimes E^{\prime}\otimes D\\
&-(A^{\prime}\otimes B+A\otimes B^{\prime})\otimes E^{\prime}\otimes D^{\prime}]\}\\
&=2\cos\alpha(\langle \mathbf{e_{0}},T_{a^{\prime}b^{\prime}}\mathbf{d^{\prime}}\rangle-\langle\mathbf{e_{0}},T_{ab}\mathbf{d^{\prime}}\rangle\\
&-\langle\mathbf{e_{0}},T_{a^{\prime}b}\mathbf{d}\rangle-\langle\mathbf{e_{0}},T_{ab^{\prime}}\mathbf{d}\rangle)\\
&+2\sin\alpha(\langle \mathbf{e_{1}},T_{ab}\mathbf{d}\rangle-\langle\mathbf{e_{1}},T_{a^{\prime}b^{\prime}}\mathbf{d}\rangle\\
&-\langle\mathbf{e_{1}},T_{a^{\prime}b}\mathbf{d^{\prime}}\rangle-\langle\mathbf{e_{1}},T_{ab^{\prime}}\mathbf{d^{\prime}}\rangle)\\
&=2\cos\alpha\langle \mathbf{e_{0}},\bm{\lambda_{0}}\rangle+2\sin\alpha(\langle \mathbf{e_{1}},\bm{\lambda_{1}}\rangle,
\end{aligned}
\end{eqnarray}
where
\begin{eqnarray}
\begin{aligned}
\label{lambda0}
\bm{\lambda_{0}}=T_{a^{\prime}b^{\prime}}\mathbf{d^{\prime}}
-T_{ab}\mathbf{d^{\prime}}-T_{a^{\prime}b}\mathbf{d}-T_{ab^{\prime}}\mathbf{d},\\
\label{lambda1}
\bm{\lambda_{1}}=T_{ab}\mathbf{d}-T_{a^{\prime}b^{\prime}}\mathbf{d}
-T_{a^{\prime}b}\mathbf{d^{\prime}}-T_{ab^{\prime}}\mathbf{d^{\prime}}.
\end{aligned}
\end{eqnarray}
In a way similar to the one used in~\cite{Su2018}, we obtain the following propositions.

\begin{proposition}
\label{Pr1}
Let $\rho$ be the density operator of a four-qubit state with correlation matrices $T_{ij}$, $i,j=1,2,3$. Then
\begin{eqnarray}
\begin{aligned}
S(\rho)=2\sqrt{F(T)},
\end{aligned}
\end{eqnarray}
where
\begin{eqnarray*}
\begin{aligned}
F(T)&=\mathop{\max}\limits_{\mathbf{a},\mathbf{a^{\prime}},\mathbf{b},\mathbf{b^{\prime}},\mathbf{d},\mathbf{d^{\prime}}}
\frac{1}{2}\left[\|\bm{\lambda_{0}}\|^{2}+\|\bm{\lambda_{1}}\|^{2}\right.\\
&\left.+\sqrt{(\|\bm{\lambda_{0}}\|^{2}+\|\bm{\lambda_{1}}\|^{2})^{2}-4\langle \bm{\lambda_{0}},\bm{\lambda_{1}}\rangle^{2}}\right]
\end{aligned}
\end{eqnarray*}
with the maximum taking over all possible measurements given by $\mathbf{a},\mathbf{a^{\prime}},\mathbf{b},\mathbf{b^{\prime}},\mathbf{d},\mathbf{d^{\prime}}$.
\end{proposition}

\begin{proposition}
\label{Pr2}
Let $\rho$ be the density operator of a four-qubit system with correlation matrices $T_{ij}$, $i,j=1,2,3$. Then
\begin{eqnarray}
\begin{aligned}
S(\rho)\leq\mathop{\max}\limits_{\mathbf{a},\mathbf{a^{\prime}},\mathbf{b},
\mathbf{b^{\prime}},\mathbf{d},\mathbf{d^{\prime}}}
2\sqrt{\|\bm{\lambda_{0}}\|^{2}+\|\bm{\lambda_{1}}\|^{2}}.
\end{aligned}
\end{eqnarray}
Furthermore, the equality is attained if the maximum is obtained for some
$\bm{\lambda_{0}}$ and $\bm{\lambda_{1}}$ such that $\bm{\lambda_{0}}\perp\bm{\lambda_{1}}$.
\end{proposition}

Let us consider the following X-type four-partite state,
\begin{widetext}
\begin{eqnarray}
\label{SX-type}
\rho=
\addtocounter{MaxMatrixCols}{10}
\left (
\begin{array}{cccccccccccccccc}
\rho_{1,1}   &0            &0         &0        &0        &0      &0       &0       &0        &0        &0      &0      &0     &0     &0     &\rho_{1,16}\\
0           &\rho_{2,2}    &0         &0        &0        &0      &0       &0       &0        &0        &0      &0      &0     &0     &0     &0\\
0           &0            &\rho_{3,3} &0        &0        &0      &0       &0       &0        &0        &0      &0      &0     &0     &0     &0\\
0           &0       &0        &\rho_{4,4}      &0        &0      &0       &0       &0        &0        &0      &0      &0     &0     &0     &0\\
0           &0       &0     &0        &\rho_{5,5}         &0      &0       &0       &0        &0        &0      &0      &0     &0     &0     &0\\
0           &0       &0     &0         &0        &\rho_{6,6}      &0       &0       &0        &0        &0      &0      &0     &0     &0     &0\\
0           &0       &0     &0         &0      &0       &\rho_{7,7}        &0       &0        &0        &0      &0      &0     &0     &0     &0\\
0           &0       &0     &0         &0      &0       &0       &\rho_{8,8}        &0        &0        &0      &0      &0     &0     &0     &0\\
0           &0       &0     &0         &0      &0       &0       &0       &\rho_{9,9}         &0        &0      &0      &0     &0     &0     &0\\
0           &0       &0     &0         &0      &0       &0       &0       &0       &\rho_{10,10}       &0      &0      &0     &0     &0     &0\\
0           &0       &0     &0         &0      &0       &0       &0       &0       &0        &\rho_{11,11}     &0      &0     &0     &0     &0\\
0           &0       &0     &0         &0      &0       &0       &0       &0       &0        &0      &\rho_{12,12}     &0     &0     &0     &0\\
0           &0       &0     &0         &0      &0       &0       &0       &0       &0        &0        &0      &\rho_{13,13}  &0     &0     &0\\
0           &0       &0     &0         &0      &0       &0       &0       &0       &0        &0        &0      &0     &\rho_{14,14}  &0     &0\\
0           &0       &0     &0         &0      &0       &0       &0       &0       &0        &0        &0      &0     &0     &\rho_{15,15}  &0\\
\rho_{16,1}           &0       &0     &0         &0      &0       &0       &0       &0       &0        &0        &0      &0     &0     &0   &\rho_{16,16}\\
\end{array}
\right ),
\end{eqnarray}
\end{widetext}
where $\rho_{1,1}+\rho_{2,2}+\rho_{3,3}+\rho_{4,4}+\rho_{5,5}+\rho_{6,6}+\rho_{7,7}+\rho_{8,8}+\rho_{9,9}+\rho_{10,10}+\rho_{11,11}+\rho_{12,12}+\rho_{13,13}+\rho_{14,14}+\rho_{15,15}+\rho_{16,16}=1$ and $\rho_{1,1}\rho_{16,16}\geq|\rho_{1,16}|^{2}$ $(\rho_{1,16}=\rho_{16,1})$.
We have, see Appendix A,
\begin{eqnarray}
\begin{aligned}
\label{eq-nonlocality}
S(\rho)=max\left\{16\sqrt{2}|\rho_{1,16}|,4\sqrt{2}|N|\right\},
\end{aligned}
\end{eqnarray}
where
$N=\rho_{1,1}-\rho_{2,2}-\rho_{3,3}+\rho_{4,4}-\rho_{5,5}+\rho_{6,6}+\rho_{7,7}-\rho_{8,8}-\rho_{9,9}+\rho_{10,10}+\rho_{11,11}-\rho_{12,12}+\rho_{13,13}-\rho_{14,14}-\rho_{15,15}+\rho_{16,16}$.
In particular, when $\rho_{1,1}=\rho_{16,16}=\rho_{1,16}=\rho_{16,1}=\frac{1}{2}$, $\rho$ reduces to the GHZ state $|GHZ\rangle$ and one obtains $S(\rho)=8\sqrt{2}$, which is the same result as the one given in~\cite{Seevinck2002}.

Similarly, for the following X-type four qubit state,
\begin{widetext}
\begin{eqnarray}
\label{S4-type}
\rho'=
\addtocounter{MaxMatrixCols}{10}
\left (
\begin{array}{cccccccccccccccc}
\rho_{1,1}   &0            &0         &0        &0        &0      &0       &0       &0        &0        &0      &0      &0     &0     &0     &\rho_{1,16}\\
0           &\rho_{2,2}    &0         &0        &0        &0      &0       &0       &0        &0        &0      &0      &0     &0     &\rho_{2,15}     &0\\
0           &0            &\rho_{3,3} &0        &0        &0      &0       &0       &0        &0        &0      &0      &0     &\rho_{3,14}     &0     &0\\
0           &0       &0        &\rho_{4,4}      &0        &0      &0       &0       &0        &0        &0      &0      &\rho_{4,13}     &0     &0     &0\\
0           &0       &0     &0        &\rho_{5,5}         &0      &0       &0       &0        &0        &0      &\rho_{5,12}      &0     &0     &0     &0\\
0           &0       &0     &0         &0        &\rho_{6,6}      &0       &0       &0        &0        &\rho_{6,11}      &0      &0     &0     &0     &0\\
0           &0       &0     &0         &0      &0       &\rho_{7,7}        &0       &0        &\rho_{7,10}        &0      &0      &0     &0     &0     &0\\
0           &0       &0     &0         &0      &0       &0       &\rho_{8,8}        &\rho_{8,9}        &0        &0      &0      &0     &0     &0     &0\\
0           &0       &0     &0         &0      &0       &0       &\rho_{9,8}       &\rho_{9,9}         &0        &0      &0      &0     &0     &0     &0\\
0           &0       &0     &0         &0      &0       &\rho_{10,7}       &0       &0       &\rho_{10,10}       &0      &0      &0     &0     &0     &0\\
0           &0       &0     &0         &0      &\rho_{11,16}       &0       &0       &0       &0        &\rho_{11,11}     &0      &0     &0     &0     &0\\
0           &0       &0     &0         &\rho_{12,5}      &0       &0       &0       &0       &0        &0      &\rho_{12,12}     &0     &0     &0     &0\\
0           &0       &0     &\rho_{13,4}         &0      &0       &0       &0       &0       &0        &0        &0      &\rho_{13,13}  &0     &0     &0\\
0           &0       &\rho_{14,3}     &0         &0      &0       &0       &0       &0       &0        &0        &0      &0     &\rho_{14,14}  &0     &0\\
0           &\rho_{15,2}       &0     &0         &0      &0       &0       &0       &0       &0        &0        &0      &0     &0     &\rho_{15,15}  &0\\
\rho_{16,1}           &0       &0     &0         &0      &0       &0       &0       &0       &0        &0        &0      &0     &0     &0   &\rho_{16,16}
\end{array}
\right),
\end{eqnarray}
\end{widetext}
where all the anti-diagonal entities are zero except for two nonzero entities $\rho_{i,j}$ and $\rho_{j,i}$ for any $i,j$ with $i+j=17$. We have
\begin{eqnarray}
\begin{aligned}
\label{eq-nonlocalitys}
S(\rho')=max\left\{16\sqrt{2}|\rho_{i,j}|,4\sqrt{2}|N|\right\},
\end{aligned}
\end{eqnarray}
where
$N=\rho_{1,1}-\rho_{2,2}-\rho_{3,3}+\rho_{4,4}-\rho_{5,5}+\rho_{6,6}+\rho_{7,7}-\rho_{8,8}
-\rho_{9,9}+\rho_{10,10}+\rho_{11,11}-\rho_{12,12}+\rho_{13,13}-\rho_{14,14}-\rho_{15,15}+\rho_{16,16}$.

\section{GENUINELY ACCESSIBLE AND INACCESSIBLE FOUR-PARTITE NONLOCALITY IN SCHWARZSCHILD BLACK HOLE}

Now we consider the genuinely accessible and inaccessible four-partite nonlocalities in Schwarzschild black hole.
In~\cite{Phy2024}, the authors quantized the Dirac field in the Schwarzschild black hole. The Kruskal vacuum and excited states
in the Schwarzschild spacetime can be written as ~\cite{Wang2010,Wu2022,Phy2024},
\begin{eqnarray}
\begin{aligned}
\label{eqQ}
&|0\rangle_{K}=\frac{1}{\sqrt{e^{-\frac{\omega}{T}}+1}}|0\rangle_{out}|0\rangle_{in}+\frac{1}{\sqrt{e^{\frac{\omega}{T}}+1}}|1\rangle_{out}|1\rangle_{in},\\
&|1\rangle_{K}=|1\rangle_{out}|0\rangle_{in},
\end{aligned}
\end{eqnarray}
where $T=\frac{1}{8\pi M}$ ($M$ is the mass of the Schwarzschild black hole) is the Hawking temperature, $|n\rangle_{out}$ and $|n\rangle_{in}$ correspond to the fermionic modes outside the event horizon and the antifermionic modes inside the event horizon, respectively. Note that the transition from Kruskal modes to Schwarzschild modes in Eqs.(\ref{eqQ}) is the two-mode squeezing transformations.

At first, the four observers share the following four-partite GHZ-state at the asymptotically flat region of the Schwarzschild black hole,
\begin{eqnarray}
\begin{aligned}
\label{eq-4qubits}
|\psi\rangle_{1,\ldots,4}=\alpha|0000\rangle+\sqrt{1-\alpha^{2}}|1111\rangle,
\end{aligned}
\end{eqnarray}
where $0\leq\alpha\leq 1$. Suppose that there are $n$ $(n<4)$ observers hovering near the event horizon of the black hole, while $4-n$ observers remain at the asymptotically flat region. That is, there are $n$ observers inside the event horizon of the black hole due to the Hawking effect.
From Eqs.(\ref{eqQ}), under the two-mode squeezing transformations, Eq.(\ref{eq-4qubits}) can be written as
\begin{eqnarray*}
\begin{aligned}
|\psi\rangle_{1,\ldots,4+n}&=\alpha\bigg[|\overline{0}\rangle\bigotimes^{n}_{i=1}\bigg(\frac{1}{\sqrt{e^{-\frac{\omega}{T}}+1}}|0\rangle_{out,i}|0\rangle_{in,i}\\
&+\frac{1}{\sqrt{e^{\frac{\omega}{T}}+1}}|1\rangle_{out}|1\rangle_{in}\bigg)\bigg]\\
&+\sqrt{1-\alpha^{2}}\bigg[|\overline{1}\rangle\bigotimes^{n}_{i=1}(|1\rangle_{out}|1\rangle_{in})\bigg],
\end{aligned}
\end{eqnarray*}
where $|\overline{0}\rangle=|0\rangle_{n+1}|0\rangle_{n+2}\ldots|0\rangle_{4}$ and $|\overline{1}\rangle=|1\rangle_{n+1}|1\rangle_{n+2}\cdots|1\rangle_{4}$.

By tracing over physically accessible $n-p$ modes and inaccessible $n-q$ modes of $|\psi\rangle_{1,\ldots,4+n}$, we obtain state $\rho_{4-n,p,q}$ $(p+q=n)$ consisting of $4-n$ Kruskal modes, $p$ accessible modes for the exterior region and $q$ inaccessible modes for the interior region of the black hole.
The density operator $\rho_{4-n,p,q}$ has the following \cite{Phy2024},
\begin{eqnarray}
\label{ex1}
\rho_{4-n,p,q}=
\left (
\begin{array}{cc}
M            &V\\
V^{T}        &N\\
\end{array}
\right ),
\end{eqnarray}
where $M$, $V$ and $N$ are $8\times8$ matrices, see Appendix B\ref{App2} for the detailed expressions.

From Eq.(\ref{S4-type}), we get the genuine four-partite nonlocality among $4-n$ Kruskal modes, $p$ accessible modes and $q$ inaccessible modes as follows,
\begin{eqnarray}
\small
\begin{aligned}
\label{eq-4nonlocalitys}
S(\rho_{4-n,p,q})=max\left\{\frac{16\sqrt{2}\alpha\sqrt{1-\alpha^{2}}}{\sqrt{(e^{-\frac{\omega}{T}}+1)^{p}(e^{\frac{\omega}{T}}+1)^{q}}},4\sqrt{2}|N|\right\},
\end{aligned}
\end{eqnarray}
where
$N=\rho_{1,1}-\rho_{2,2}-\rho_{3,3}+\rho_{4,4}-\rho_{5,5}+\rho_{6,6}+\rho_{7,7}-\rho_{8,8}-\rho_{9,9}+\rho_{10,10}+\rho_{11,11}-\rho_{12,12}
+\rho_{13,13}-\rho_{14,14}-\rho_{15,15}+\rho_{16,16}$.
From Eq.(\ref{eq-4nonlocalitys}), we observe that the genuine four-partite nonlocality depends not only the initial state parameters $\alpha$ but also $T$, $p$ and $q$.

\begin{figure}[t]
\centering
\includegraphics[scale=0.55]{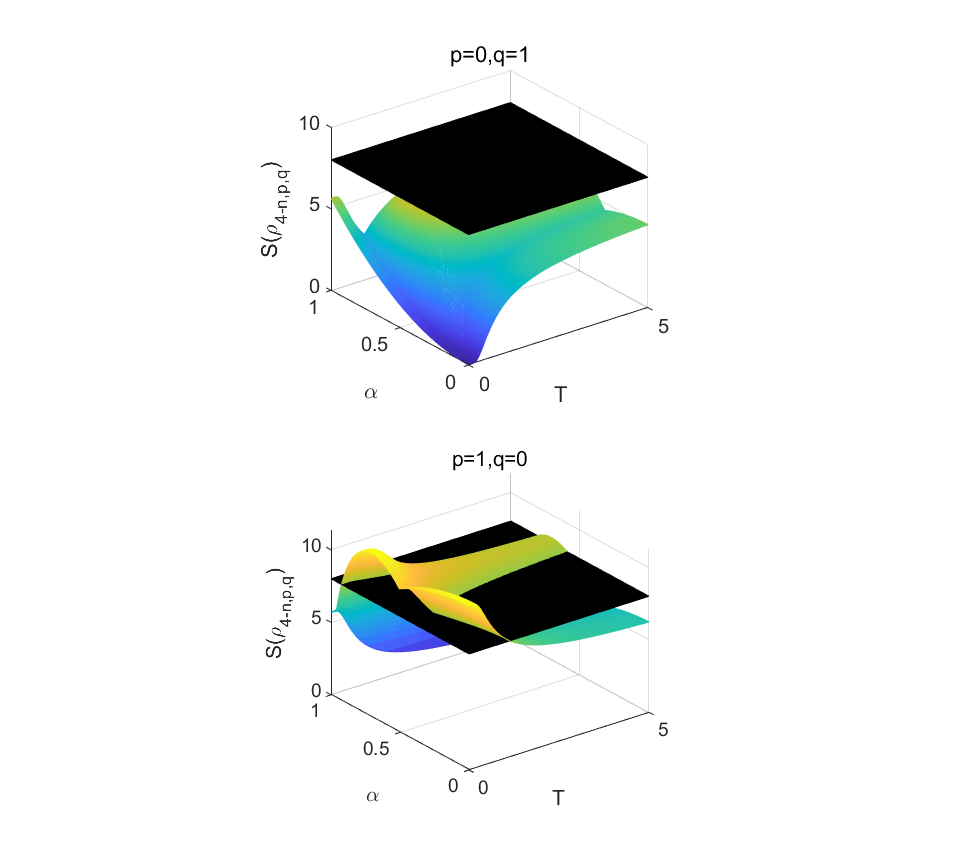}
\caption{The genuine four-partite nonlocality $S(\rho_{4-n,p,q})$ $(n=1)$ as functions of the Hawking temperature $T$ and initial state parameter $\alpha$ for
different $p$, $q$, where $\omega=1$. The black plane is $S(\rho_{4-n,p,q})=8$.}
\label{Fig2}
\end{figure}

\begin{figure}[t]
\centering
\includegraphics[scale=0.55]{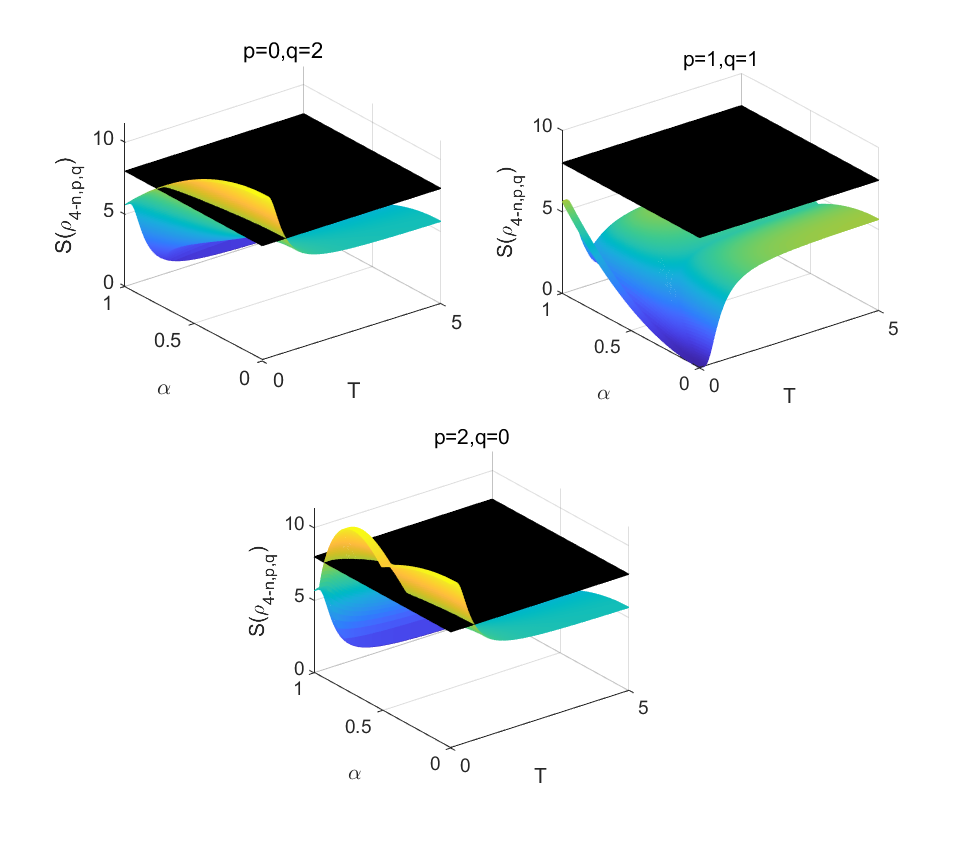}
\caption{The genuine four-partite nonlocality $S(\rho_{4-n,p,q})$ $(n=2)$ as functions of the Hawking temperature $T$ and initial state parameter $\alpha$ for
different $p$, $q$, where $\omega=1$. The black plane is $S(\rho_{4-n,p,q})=8$.}
\label{Fig3}
\end{figure}

\begin{figure}[h]
\centering
\includegraphics[scale=0.55]{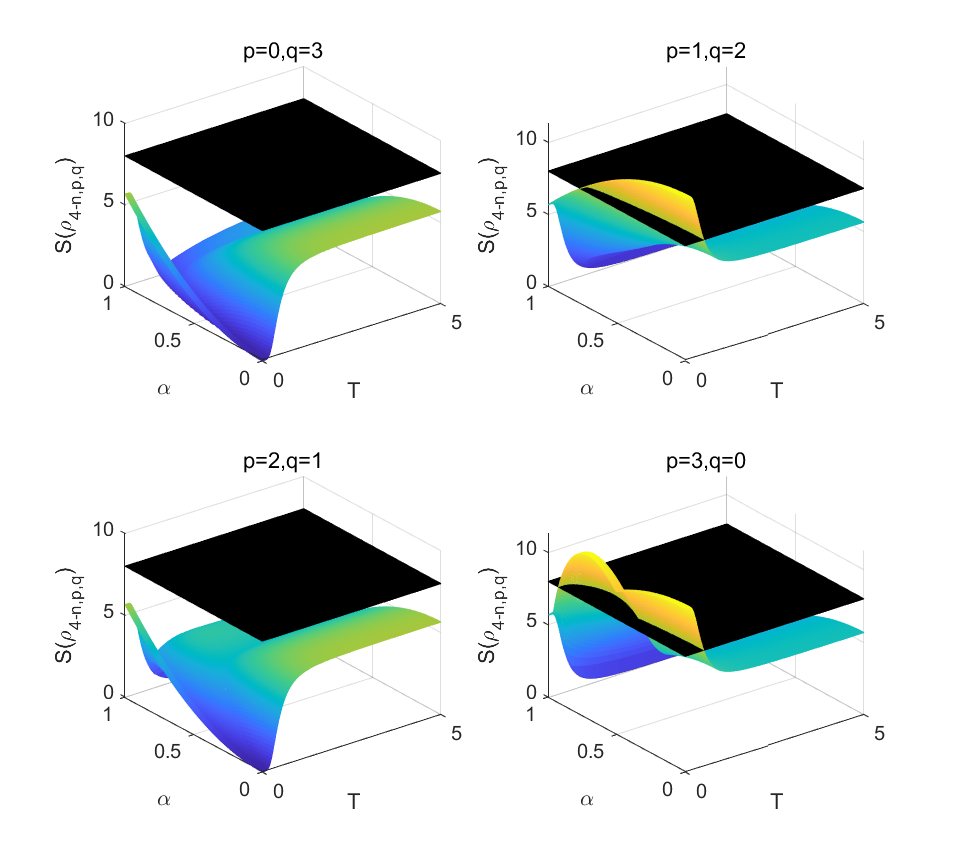}
\caption{The genuine four-partite nonlocality $S(\rho_{4-n,p,q})$ $(n=3)$ as functions of the Hawking temperature $T$ and initial state parameter $\alpha$ for
different $p$, $q$, where $\omega=1$. The black plane is $S(\rho_{4-n,p,q})=8$.}
\label{Fig4}
\end{figure}

In Fig.\ref{Fig2}-\ref{Fig4}, we depict the genuine four-partite nonlocality $S(\rho_{4-n,p,q})$ as functions of the Hawking temperature $T$ and the initial state parameter $\alpha$ for different $p$ and $q$. Here, $n=1,2,3$ since $n<4$. There are 9 possible scenarios for the genuine four-partite nonlocality $S(\rho_{4-n,p,q})$ in Fig.\ref{Fig2}-\ref{Fig4}. Note that the nonlocality $S(\rho_{4-n,p,q})$ encompasses both all accessible and inaccessible nonlocalities. That is, the genuine four-partite nonlocality $S(\rho_{4-n,p,q})$ is an accessible nonlocality for $p=n$, while an inaccessible nonlocality for $p\neq n$.

From Fig.\ref{Fig2}-\ref{Fig4}, we see that when $p=n$, $S(\rho_{4-n,p,q})$ monotonically decreases with the increase of the Hawking temperature $T$ for fixed $\alpha$. On the other hand, if the value of $T$ is fixed, $S(\rho_{4-n,p,q})$ is non-monotonic with the increase of the initial state parameter $\alpha$, see the figures for $p=1,q=0$, $p=2,q=0$ and $p=3,q=0$. In addition, when $p\neq n$, if the value of $\alpha$ is fixed, $S(\rho_{4-n,p,q})$ is non-monotonic with the increase of the Hawking temperature $T$. In particular, when $\alpha=0$ ($\alpha=1$), $S(\rho_{4-n,p,q})$ is monotonically increasing (decreasing) with the increase of the Hawking temperature $T$. On the contrary, if the value of $T$ is fixed, $S(\rho_{4-n,p,q})$ is non-monotonic with the increase of the initial state parameter $\alpha$. When $T\rightarrow0$, $S(\rho_{4-n,p,q})$ monotonically increases with the increase of initial state parameter $\alpha$, see figures for $p=0,q=1$, $p=1,q=1$, $p=0,q=3$ and $p=2,q=1$. Therefore, we conclude that the Hawking effect of the black hole destroys the accessible genuine four-partite nonlocality, while having both negative and positive impacts on the inaccessible genuine four-partite nonlocality. 
Additionally, regarding the genuine N-partite entanglement in Schwarzschild black hole, we know that the Hawking effect reduces the accessible genuine N-partite entanglement and enhances monotonically or non-monotonically the inaccessible genuine N-partite entanglement from \cite{Phy2024}. Thus, the Hawking effect of the black hole undermines the accessible genuine four-partite nonlocality and genuine N-partite entanglement, while producing both adverse and favorable influences on the inaccessible genuine four-partite nonlocality and genuine N-partite entanglement.

Interestingly, we find that the value of $S(\rho_{4-n,p,q})$ is always smaller than $8$ with the increase of Hawking temperature $T$ and the decrease of $\alpha$ when $p=0,q=1$, $p=1,q=1$, $p=0,q=3$ and $p=2,q=1$. Thus, the physically inaccessible genuine four-partite nonlocality cannot be generated in these scenario.
However, we find that $S(\rho_{4-n,p,q})$ has regions both smaller than $8$ and larger than $8$ with the increase of Hawking temperature $T$ and the decrease of $\alpha$ when $p=0,q=2$ and $p=1,q=2$. That is, the Hawking effect can generate the physically inaccessible genuine four-partite nonlocality for fermion fields in Schwarzschild black hole.
This result indicates that the genuine four-partite nonlocality can pass through the event horizon of black hole.

\section{GENUINELY four-PARTITE NONLOCALITY IN SCHWARZSCHILD-DE SITTER BLACK HOLE}
In this section, we study the genuine four-partite nonlocality in the Schwarzschild-de Sitter (SdS) black hole spacetime, distinguished by the existence of a black hole event horizon (BEH) and a cosmological event horizon (CEH). The $r_{H}$ (BEH) and $r_{C}$ (CEH) can be written as \cite{Wu2024},
\begin{eqnarray}
\begin{aligned}
\label{eq-R}
&r_{H}=\frac{2}{\sqrt{\Lambda}}\cos\left[\frac{\pi+\arccos(3M\sqrt{\Lambda})}{3}\right],\\
&r_{C}=\frac{2}{\sqrt{\Lambda}}\cos\left[\frac{\arccos(3M\sqrt{\Lambda})-\pi}{3}\right],
\end{aligned}
\end{eqnarray}
where $\Lambda$ is a cosmological constant in $(3+1)$-spacetime dimensions~\cite{Kottler1918}. The surface gravities of the black hole and the expanding universe are given by
\begin{eqnarray}
\begin{aligned}
\label{eq-Rr}
&k_{H}=\frac{\Lambda(2r_{H}+r_{C})(r_{C}-r_{H})}{6r_{H}},\\
-&k_{C}=\frac{\Lambda(2r_{C}+r_{H})(r_{H}-r_{C})}{6r_{C}},
\end{aligned}
\end{eqnarray}
respectively. The Hawking temperature of the black hole $T_{H}=\frac{k_{H}}{2\pi}$ is greater than the Hawking temperature of the expanding universe $T_{C}=\frac{k_{C}}{2\pi}$~\cite{Bhattacharya.2013,Bhattacharya2022,Chowdhury2023}.

\begin{figure}[h]
\centering
\includegraphics[scale=0.42]{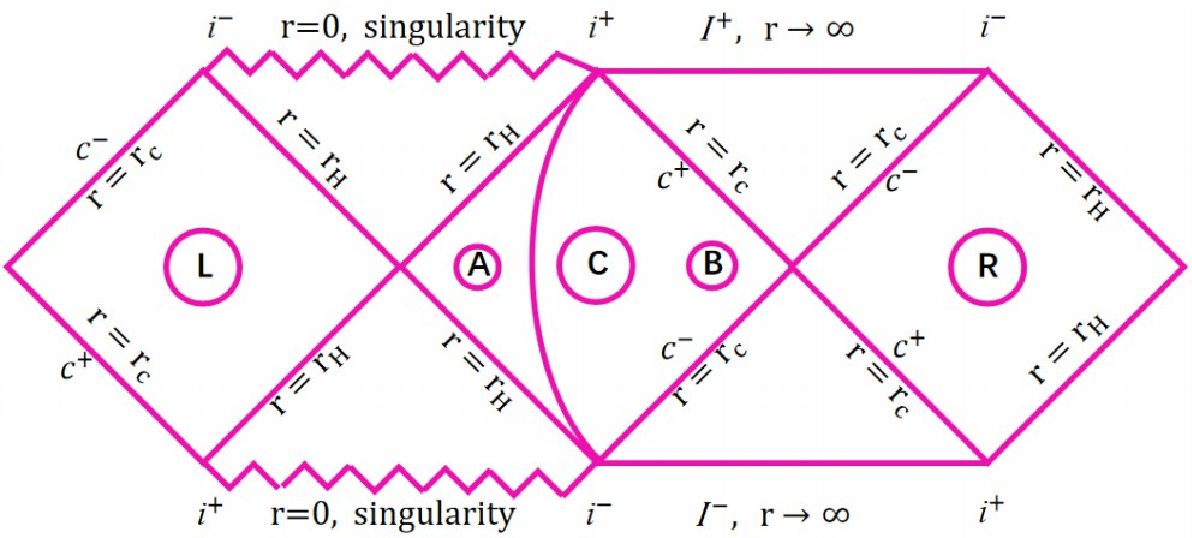}
\caption{The SdS spacetime with thermal opaque membrane~\cite{Wu2024}.}
\label{Fig1}
\end{figure}

In~\cite{Wu2024} the authors use the thermally opaque membrane to divide the $C$ region into two sub-regions, that is, $A$ and $B$ $(C=A\cup B)$, see Fig.\ref{Fig1}.
It is found that the $n$ observers positioned at the BEH can perceive the Hawking radiation at temperature $T_{H}$, whereas the $m$ observers positioned at the CEH can perceive the Hawking radiation at temperature $T_{C}$.

Therefore, the Kruskal vacuum state and the excited state in the black hole spacetime can be written as~\cite{Wu2024},
\begin{eqnarray}
\begin{aligned}
\label{eqH}
&|0\rangle_{k_{H}}=\cos r|0_{A},0_{L}\rangle+\sin r|1_{A},1_{L}\rangle,\\
&|1\rangle_{k_{H}}=|1_{A},0_{L}\rangle,
\end{aligned}
\end{eqnarray}
where $\cos r=\frac{1}{\sqrt{e^{-\frac{\omega}{T_{H}}}+1}}$, $|n_{A}\rangle$ and $|n_{L}\rangle$ represent the number states of fermion outside the event horizon and the antifermion inside the event horizon of the black hole, respectively. Correspondingly, the Kruskal vacuum state and the excited state in the expanding universe can be denoted by~\cite{Wu2024}
\begin{eqnarray}
\begin{aligned}
\label{eqE}
&|0\rangle_{k_{C}}=\cos\omega|0_{B},0_{R}\rangle+\sin\omega|1_{B},1_{R}\rangle,\\
&|1\rangle_{k_{C}}=|1_{B},0_{R}\rangle,
\end{aligned}
\end{eqnarray}
where $\cos\omega=\frac{1}{\sqrt{e^{-\frac{\omega}{T_{C}}}+1}}$.

We first examine a four-partite GHZ-state that consists of $n$ Kruskal modes $k_{H}$ and $m$ Kruskal modes $k_{C}$,
\begin{eqnarray}
\begin{aligned}
\label{eq-44qubit}
|\psi\rangle_{1,\ldots,4}&=\alpha|0^{1}_{k_{H}},\ldots,0^{n}_{k_{H}},0^{1}_{k_{C}},\ldots,0^{m}_{k_{C}}\rangle\\
&+\sqrt{1-\alpha^{2}}|1^{1}_{k_{H}},\ldots,1^{n}_{k_{H}},1^{1}_{k_{C}},\ldots,1^{m}_{k_{C}}\rangle,
\end{aligned}
\end{eqnarray}
where $n$ $(0<n<N)$, $m$ $(0<m<N)$ and $n+m=4$.
We assign the $n$ modes to the BEH in sub-region $A$ of $C$, while the $m$ modes are situated at the CEH in the sub-region $B$ of $C$. As a result of the Hawking effects in the SdS spacetime, the region $L$ in Fig.\ref{Fig1} experiences the emergence of the additional $n$ modes, while the region $R$ witnesses the appearance of the $m$ modes.
From Eqs.(\ref{eqH}) and (\ref{eqE}), the Eq.(\ref{eq-44qubit}) can be rewritten as
\begin{eqnarray}
\begin{aligned}
|\psi\rangle_{1,\ldots,8}&=\alpha\left[\bigotimes^{n}_{i=1}(\cos r|0^{i}_{A},0^{i}_{L}\rangle+\sin r|1^{i}_{A},1^{i}_{L}\rangle)\right.\\
&\left.\bigotimes^{m}_{j=1}(\cos\omega|0^{j}_{B},0^{j}_{R}\rangle+\sin\omega|1^{j}_{B},1^{j}_{R}\rangle)\right]\\
&+\sqrt{1-\alpha^{2}}\left[\bigotimes^{n}_{u=1}(|1^{u}_{A},0^{u}_{L}\rangle)\bigotimes^{m}_{v=1}(|1^{v}_{B},0^{v}_{R}\rangle)\right].
\end{aligned}
\end{eqnarray}

By tracing over the physically inaccessible modes in the sub-regions $L$ and $R$, from~\cite{Wu2024} the density operator $\rho_{4}$ can be expressed as
\begin{eqnarray}
\rho_{4}=
\left (
\begin{array}{cc}
M_{A}            & M_{X}\\
M_{X}^{T}        & M_{B}\\
\end{array}
\right )
\end{eqnarray}
in the $16$ bases
$\left\{|0^{1}_{A},\ldots,0^{n}_{A},0^{1}_{B},\ldots,0^{m}_{B}\rangle,|0^{1}_{A},\ldots,0^{n}_{A},0^{1}_{B},\right.\\
\ldots,0^{m-1}_{B},0^{m}_{B}\rangle,
\left.\ldots,|1^{1}_{A},\ldots,1^{n}_{A},1^{1}_{B},\ldots,1^{m-1}_{B},0^{m}_{B}\rangle,|1^{1}_{A},\right.\\
\left.\ldots,1^{n}_{A},1^{1}_{B},\ldots,
1^{m}_{B}\rangle\right\}$,
where the basis corresponding to the element $\alpha^{2}\cos^{2(n-k)}r\sin^{2k}r\cos^{2(m-l)}\omega\sin^{2l}\omega$ contains $k$ $``1_{A}"$ and $l$ $``1_{B}"$.
The sub-matrices $M_{A}$, $M_{B}$ and $M_{X}$ are detailed in Appendix C\ref{App3}.

Therefore, according to Eq.(\ref{eq-nonlocality}), we get the genuine four-partite nonlocality in the SdS black hole spacetime as follows,
\begin{eqnarray}
\begin{aligned}
\label{eq-4nonlocality}
S(\rho_{4})=max\left\{16\sqrt{2}\alpha\sqrt{1-\alpha^{2}}|\cos^{n}r\cos^{m}\omega|,4\sqrt{2}|N|\right\},
\end{aligned}
\end{eqnarray}
where
$N=\rho_{1,1}-\rho_{2,2}-\rho_{3,3}+\rho_{4,4}-\rho_{5,5}+\rho_{6,6}+\rho_{7,7}-\rho_{8,8}-\rho_{9,9}+\rho_{10,10}+\rho_{11,11}-\rho_{12,12}
+\rho_{13,13}-\rho_{14,14}-\rho_{15,15}+\rho_{16,16}$.
From Eq.(\ref{eq-4nonlocality}), it is apparent that the genuine four-partite nonlocality relies on not only just the initial parameters $\alpha$, $n$ and $m$, but also the mass $M$ of the black hole and the cosmological constant $\Lambda$.

\begin{figure}[h]
\centering
\includegraphics[scale=0.55]{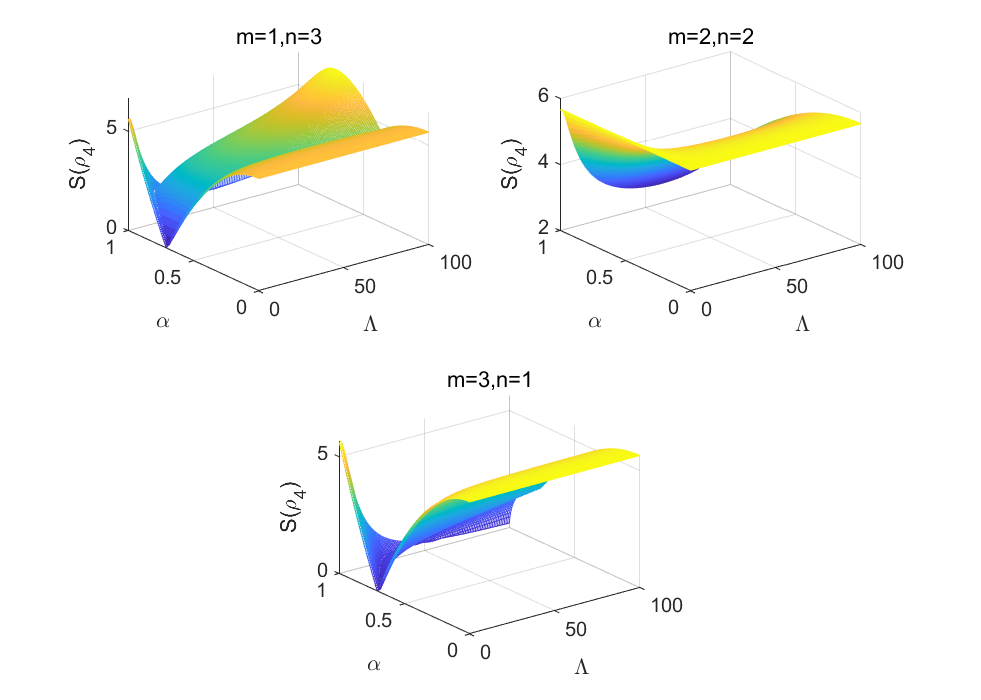}
\caption{The genuine four-partite nonlocality $S(\rho_{4})$ as functions of the cosmological constant $\Lambda$ and the initial parameter $\alpha$ for
different $n$ and $m$, where $M=0.033$ and $\omega=1$.}
\label{Fig5}
\end{figure}

In Fig.\ref{Fig5}, we depict the $S(\rho_{4})$ (the representative element of genuine four-partite nonlocality) as functions of the cosmological constant $\Lambda$ and the initial parameter $\alpha$ for
different $n$ and $m$. From Fig.\ref{Fig5}, we observe that $S(\rho_{4})=4\sqrt{2}$ when $\alpha=0$. Besides, when $\alpha=1$, $S(\rho_{4})$ is non-monotonic for $m=1,n=3$ and $m=2,n=2$, and is monotonically decreasing for $m=3,n=1$. For $m=1,n=3$ and $m=3,n=1$, $S(\rho_{4})$ is non-monotonic with the increase of the cosmological constant $\Lambda$ and initial parameter $\alpha$. For $m=2,n=2$, $S(\rho_{4})=4\sqrt{2}$ when $\alpha=0$ or $\Lambda=0$. If the value of $\Lambda$ is fixed, $S(\rho_{4})$ is monotonically decreasing with the increase of the initial parameter $\alpha$. On the other hand, if the value of $\alpha$ is fixed, $S(\rho_{4})$ is non-monotonic with the increase of the cosmological constant $\Lambda$.

\begin{figure}[h]
\centering
\includegraphics[scale=0.55]{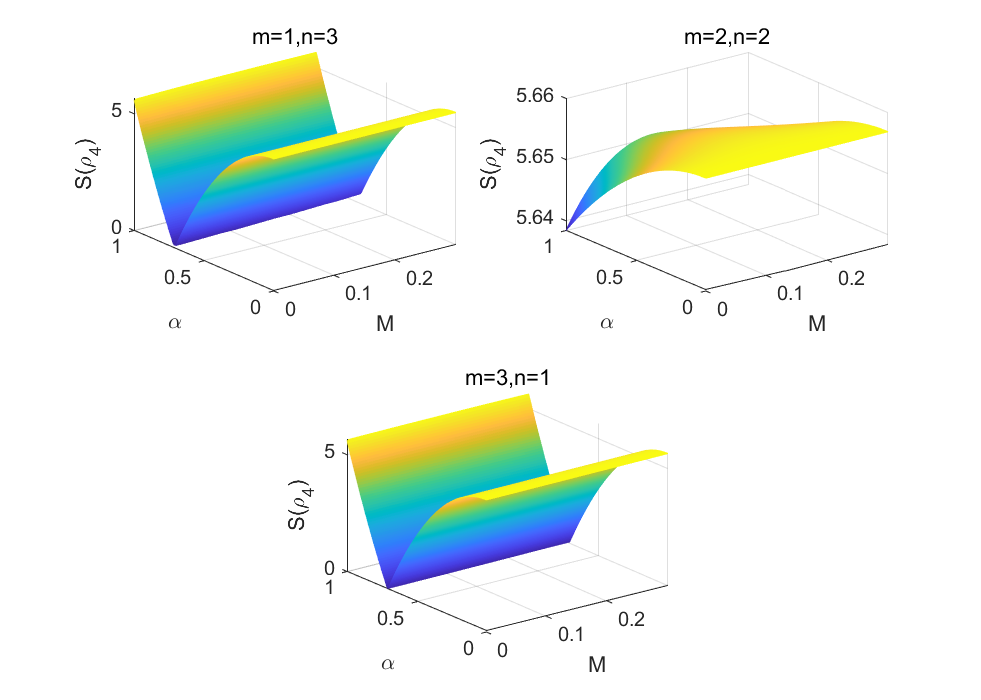}
\caption{The genuine four-partite nonlocality $S(\rho_{4})$ as functions of the mass $M$ and initial parameter $\alpha$ for
different $n$ and $m$, where $\Lambda=\omega=1$.}
\label{Fig6}
\end{figure}

In Fig.\ref{Fig6}, we depict $S(\rho_{4})$ as functions of the mass $M$ and initial parameter $\alpha$ for
different $n$ and $m$. Obviously, it is seen that $S(\rho_{4})=4\sqrt{2}$ when $\alpha=0$.
From Fig.\ref{Fig6}, $S(\rho_{4})$ is non-monotonic with the increase of the mass $M$ and initial parameter $\alpha$. Note that when $m=2,n=2$, $S(\rho_{4})$ is non-monotonic with the increase of $M$ if the value of $\alpha$ is fixed. However, if $M$ is fixed, $S(\rho_{4})$ is monotonically increasing with increase of $\alpha$.

Therefore, from Fig.\ref{Fig5} and Fig.\ref{Fig6} we have that for different modes, the $S(\rho_{4})$ has different properties depending on different mass $M$ and cosmological constant $\Lambda$. The Hawking effect of the SdS spacetime may either disrupt or enhance the four-partite nonlocality. However, $S(\rho_{4})$ is maximized at $4\sqrt{2}$, and it does not show genuine four-partite nonlocality.

 Furthermore, regarding the genuine N-partite entanglement in Schwarzschild-de Sitter black hole, we know that the Hawking effect of the black hole destroys quantum entanglement and the Hawking effect of the expanding universe can enhance quantum entanglement\cite{Wu2024}. Through our research, the Hawking effect of the black hole may enhance the $S(\rho_{4})$. This result is helpful for us to understand the Hawking effect in single-event horizon spacetime.

\section{Conclusion}
We have investigated the genuine four-partite nonlocality in the background of the Schwarzschild black hole and the SdS spacetime.
Firstly, we have presented quantitative analysis on the genuine four-partite nonlocality and obtained an analytical expression of the genuine four-partite nonlocality.
Subsequently, we have studied the genuinely accessible and inaccessible four-partite nonlocality in Schwarzschild black hole. We initially supposed that four observes share a four-partite GHZ-state in the asymptotically flat region. Then, let $n$ $(n<4)$ observers hover near the event horizon of the black hole, while the $4-n$ observers remain at the asymptotically flat region. We have derived analytical expressions for both physically accessible and inaccessible genuine four-partite nonlocalities. It has been illustrated that the Hawking effect of the black hole destroys the accessible genuine four-partite nonlocality, while having both negative and positive impacts on the inaccessible genuine four-partite nonlocality non-locality.
In previous studies~\cite{Eur2022,Li2022}, the Hawking effect cannot generate the physically inaccessible genuine tripartite nonlocality with pure initial states. But in ~\cite{Zhang.2023}, Zhang \emph{et al.} discovered for the first time that the Hawking effect can generate physically inaccessible genuine tripartite nonlocality with certain mixed initial states. In the paper, we show that the Hawking effect can generate the physically inaccessible genuine four-partite nonlocality with pure initial states.

Finally, we have considered the genuine four-partite nonlocality in the Schwarzschild-de Sitter black hole spacetime with the black hole event horizon and the cosmological event horizon. Let $n$ observers near BEH and $m$ $(n+m=4)$ observers near CEH. Then, we have obtained the analytical expression of the genuine four-partite nonlocality in the SdS spacetime. We conclude that for different modes, the $S(\rho_{4})$ has different properties under the impacts of the mass $M$ and the cosmological constant $\Lambda$, that is, the Hawking effect may either disrupt or enhance genuine four-partite nonlocality. These conclusions provide a more comprehensive understanding of genuine four-partite nonlocality in curved spacetime.

\begin{acknowledgments}
 This work is supported by the National Natural Science Foundation of China (NSFC) under Grant Nos. 12204137, 12075159 and 12171044; the specific research fund of the Innovation Platform for Academicians of Hainan Province under Grant No. YSPTZX202215 and Hainan Academician Workstation (Changbin Yu).
\end{acknowledgments}

\appendix

\section{APPENDIX A: GENUINE NONLOCALITY OF THE FOUR-QUBIT X-TYPE STATE}
\label{App1}
We analyze the genuine nonlocality of the four-qubit states $\rho$ by calculating the value $S(\rho)$.
The $9$ correlation matrices $T_{ij}$, $i,j=1,2,3$, are as follows,
\begin{eqnarray*}
T_{11}=(t_{11kl})&&=
\left (
\begin{array}{ccc}
\rho_{1,16}+\rho_{16,1}              &i\rho_{1,16}-i\rho_{16,1}           &0\\
i\rho_{1,16}-i\rho_{16,1}           &-\rho_{1,16}-\rho_{16,1}            &0\\
0                                    &0                                   &0\\
\end{array}
\right )\\
&&=2\rho_{1,16}
\left (
\begin{array}{ccc}
1              &0           &0\\
0              &-1          &0\\
0              &0           &0\\
\end{array}
\right ),
\end{eqnarray*}
\begin{eqnarray*}
T_{12}=(t_{12kl})&&=
\left (
\begin{array}{ccc}
i\rho_{1,16}-i\rho_{16,1}            &-\rho_{1,16}-\rho_{16,1}           &0\\
-\rho_{1,16}-\rho_{16,1}             &-i\rho_{1,16}+i\rho_{16,1}            &0\\
0                                    &0                                   &0\\
\end{array}
\right )\\
&&=2\rho_{1,16}
\left (
\begin{array}{ccc}
0              &-1           &0\\
-1              &0          &0\\
0              &0           &0\\
\end{array}
\right ),
\end{eqnarray*}
\begin{eqnarray*}
T_{22}=(t_{22kl})&&=
\left (
\begin{array}{ccc}
-\rho_{1,16}-\rho_{16,1}               &-i\rho_{1,16}+i\rho_{16,1}           &0\\
-i\rho_{1,16}+i\rho_{16,1}             &\rho_{1,16}+\rho_{16,1}              &0\\
0                                      &0                                    &0\\
\end{array}
\right )\\
&&=2\rho_{1,16}
\left (
\begin{array}{ccc}
-1              &0           &0\\
0               &1           &0\\
0               &0           &0\\
\end{array}
\right ),
\end{eqnarray*}
\begin{eqnarray*}
T_{33}=(t_{33kl})=
\left (
\begin{array}{ccc}
0               &0           &0\\
0               &0           &0\\
0               &0           &N\\
\end{array}
\right ),
\end{eqnarray*}
where $N=\rho_{1,1}-\rho_{2,2}-\rho_{3,3}+\rho_{4,4}-\rho_{5,5}+\rho_{6,6}+\rho_{7,7}-\rho_{8,8}-\rho_{9,9}+\rho_{10,10}+\rho_{11,11}-\rho_{12,12}+\rho_{13,13}-\rho_{14,14}-\rho_{15,15}+\rho_{16,16}$. Besides, $T_{21}=T_{12}$, $T_{13}=T_{23}=T_{31}=T_{32}=0$.

According to Proposition \ref{Pr2}, we need to calculate the value of $\mathop{\max}\limits_{\mathbf{a},\mathbf{a^{\prime}},\mathbf{b},\mathbf{b^{\prime}},\mathbf{d},\mathbf{d^{\prime}}}2\sqrt{\|\bm{\lambda_{0}}\|^{2}+\|\bm{\lambda_{1}}\|^{2}}$,
where the maximum is taken over all possible measurement variables $\mathbf{a},\mathbf{a^{\prime}},\mathbf{b},\mathbf{b^{\prime}},\mathbf{d},\mathbf{d^{\prime}}$.
From Eq.(\ref{lambda0}), we obtain
\begin{widetext}
\begin{eqnarray*}
\begin{aligned}
\|\bm{\lambda_{0}}\|^{2}&=(\mathbf{d^{\prime}}T_{a^{\prime}b^{\prime}}-\mathbf{d^{\prime}}T_{ab}-\mathbf{d}T_{a^{\prime}b}-\mathbf{d}T_{ab^{\prime}})(T_{a^{\prime}b^{\prime}}\mathbf{d^{\prime}}^{T}-T_{ab}\mathbf{d^{\prime}}^{T}-T_{a^{\prime}b}\mathbf{d}^{T}-T_{ab^{\prime}}\mathbf{d}^{T})\\
&=\mathbf{d^{\prime}}T_{a^{\prime}b^{\prime}}T_{a^{\prime}b^{\prime}}\mathbf{d^{\prime}}^{T}+\mathbf{d^{\prime}}T_{ab}T_{ab}\mathbf{d^{\prime}}^{T}+\mathbf{d}T_{a^{\prime}b}T_{a^{\prime}b}\mathbf{d}^{T}+\mathbf{d}T_{ab^{\prime}}T_{ab^{\prime}}\mathbf{d}^{T}\\
&-2\mathbf{d^{\prime}}T_{ab}T_{a^{\prime}b^{\prime}}\mathbf{d^{\prime}}^{T}+2\mathbf{d}T_{a^{\prime}b}T_{ab^{\prime}}\mathbf{d}^{T}-2\mathbf{d^{\prime}}T_{a^{\prime}b^{\prime}}T_{a^{\prime}b}\mathbf{d}^{T}-2\mathbf{d^{\prime}}T_{a^{\prime}b^{\prime}}T_{ab^{\prime}}\mathbf{d}^{T}\\
&+2\mathbf{d^{\prime}}T_{ab}T_{a^{\prime}b}\mathbf{d}^{T}+2\mathbf{d^{\prime}}T_{ab}T_{ab^{\prime}}\mathbf{d}^{T}
\end{aligned}
\end{eqnarray*}
\end{widetext}
and
\begin{widetext}
\begin{eqnarray*}
\begin{aligned}
\|\bm{\lambda_{1}}\|^{2}&=(\mathbf{d}T_{ab}-\mathbf{d}T_{a'b'}-\mathbf{d'}T_{a'b}-\mathbf{d'}T_{ab'})(T_{ab}\mathbf{d}-T_{a'b'}\mathbf{d}-T_{a'b}\mathbf{d'}-T_{ab'}\mathbf{d'})\\
&=\mathbf{d}T_{ab}T_{ab}\mathbf{d}^{T}+\mathbf{d}T_{a'b'}T_{a'b'}\mathbf{d}^{T}+\mathbf{d'}T_{a'b}T_{a'b}\mathbf{d'}^{T}+\mathbf{d'}T_{ab'}T_{ab'}\mathbf{d'}^{T}\\
&-2\mathbf{d}T_{ab}T_{a'b'}\mathbf{d}^{T}+2\mathbf{d'}T_{a'b}T_{ab'}\mathbf{d'}^{T}-2\mathbf{d}T_{ab}T_{a'b}\mathbf{d'}^{T}-2\mathbf{d}T_{ab}T_{ab'}\mathbf{d'}^{T}\\
&+2\mathbf{d}T_{a'b'}T_{a'b}\mathbf{d'}^{T}+2\mathbf{d}T_{a'b'}T_{ab'}\mathbf{d'}^{T}.
\end{aligned}
\end{eqnarray*}
\end{widetext}
Therefore,
\begin{widetext}
\begin{eqnarray}
\begin{aligned}
\label{eqmax}
\|\bm{\lambda_{0}}\|^{2}+\|\bm{\lambda_{1}}\|^{2}&=(\mathbf{d}+\mathbf{d'})(T_{ab}T_{ab}+T_{a'b'}T_{a'b'}+T_{a'b}T_{a'b}+T_{ab'}T_{ab'})(\mathbf{d}^{T}+\mathbf{d'}^{T})\\
&-2\mathbf{d}(T_{ab}T_{ab}+T_{a'b'}T_{a'b'}+T_{a'b}T_{a'b}+T_{ab'}T_{ab'})\mathbf{d'}^{T}\\
&+2\mathbf{d}(T_{a'b}T_{ab'}-T_{ab}T_{a'b'})\mathbf{d}^{T}+2\mathbf{d'}(T_{a'b}T_{ab'}-T_{ab}T_{a'b'})\mathbf{d'}^{T}\\
&+2\mathbf{d}(T_{a'b'}T_{a'b}+T_{a'b'}T_{ab'}-T_{ab}T_{a'b}-T_{ab}T_{ab'})\mathbf{d'}^{T}\\
&+2\mathbf{d'}(T_{ab}T_{a'b}+T_{ab}T_{ab'}-T_{a'b'}T_{a'b}-T_{a'b'}T_{ab'})\mathbf{d}^{T}\\
&=4\rho_{1,16}^{2}\left[(a_{1}^{2}+a_{2}^{2}+a_{1}^{\prime2}+a_{2}^{\prime2})(b_{1}^{2}+b_{2}^{2}+b_{1}^{\prime2}+b_{2}^{\prime2})(d_{1}^{2}+d_{2}^{2}+d_{1}^{\prime2}+d_{2}^{\prime2})\right.\\
&\left.+4(a_{1}a'_{2}-a'_{1}a_{2})(b_{1}b'_{2}-b'_{1}b_{2})(d_{1}^{2}+d_{2}^{2}+d_{1}^{\prime2}+d_{2}^{\prime2})+4(d_{1}d'_{2}-d'_{1}d_{2})(b'_{1}b_{2}-b_{1}b'_{2})(a_{1}^{2}+a_{2}^{2}+a_{1}^{\prime2}+a_{2}^{\prime2})\right.\\
&\left.+4(d_{1}d'_{2}-d'_{1}d_{2})(a'_{1}a_{2}-a_{1}a'_{2})(b_{1}^{2}+b_{2}^{2}+b_{1}^{\prime2}+b_{2}^{\prime2})\right]
+N^{2}(a_{3}^{2}+a_{3}^{\prime2})(b_{3}^{2}+b_{3}^{\prime2})(d_{3}^{2}+d_{3}^{\prime2}).
\end{aligned}
\end{eqnarray}
\end{widetext}

As $\mathbf{a},\mathbf{a'},\mathbf{b},\mathbf{b'},\mathbf{d},\mathbf{d'}\in\mathbb{R}^{3}$ are unit vectors, we have the following polar coordinate transformations
\begin{eqnarray}
\label{po1}
\left\{
\begin{aligned}
&a_{1}=\sin\alpha_{1}\sin\alpha_{2}\\
&a_{2}=\sin\alpha_{1}\cos\alpha_{2}\\
&a_{3}=\cos\alpha_{1}
\end{aligned},
\right.
\left\{
\begin{aligned}
&a'_{1}=\sin\beta_{1}\sin\beta_{2}\\
&a'_{2}=\sin\beta_{1}\cos\beta_{2}\\
&a'_{3}=\cos\beta_{1}
\end{aligned},
\right.
\end{eqnarray}
\begin{eqnarray}
\label{po2}
\left\{
\begin{aligned}
&b_{1}=\sin\alpha_{3}\sin\alpha_{4}\\
&b_{2}=\sin\alpha_{3}\cos\alpha_{4}\\
&b_{3}=\cos\alpha_{3}
\end{aligned},
\right.
\left\{
\begin{aligned}
&b'_{1}=\sin\beta_{3}\sin\beta_{4}\\
&b'_{2}=\sin\beta_{3}\cos\beta_{4}\\
&b'_{3}=\cos\beta_{3}
\end{aligned},
\right.
\end{eqnarray}
\begin{eqnarray}
\label{po3}
\left\{
\begin{aligned}
&d_{1}=\sin\alpha_{5}\sin\alpha_{6}\\
&d_{2}=\sin\alpha_{5}\cos\alpha_{6}\\
&d_{3}=\cos\alpha_{5}
\end{aligned},
\right.
\left\{
\begin{aligned}
&d'_{1}=\sin\beta_{5}\sin\beta_{6}\\
&d'_{2}=\sin\beta_{5}\cos\beta_{6}\\
&d'_{3}=\cos\beta_{5}
\end{aligned}.
\right.
\end{eqnarray}
Substitute Eqs.(\ref{po1}), (\ref{po2}), (\ref{po3}) into the Eq.(\ref{eqmax}), we have
\begin{widetext}
\begin{eqnarray}
\begin{aligned}
\|\bm{\lambda_{0}}\|^{2}+\|\bm{\lambda_{1}}\|^{2}
&=4\rho_{1,16}^{2}\left[(\sin^{2}\alpha_{1}+\sin^{2}\beta_{1})(\sin^{2}\alpha_{3}+\sin^{2}\beta_{3})(\sin^{2}\alpha_{5}+\sin^{2}\beta_{5})\right.\\
&\left.+4\sin\alpha_{1}\sin\beta_{1}\sin\alpha_{3}\sin\beta_{3}\sin(\alpha_{2}-\beta_{2})\sin(\alpha_{4}-\beta_{4})(\sin^{2}\alpha_{5}+\sin^{2}\beta_{5})\right.\\
&\left.+4\sin\alpha_{3}\sin\beta_{3}\sin\alpha_{5}\sin\beta_{5}\sin(\beta_{4}-\alpha_{4})\sin(\alpha_{6}-\beta_{6})(\sin^{2}\alpha_{1}+\sin^{2}\beta_{1})\right.\\
&\left.+4\sin\alpha_{1}\sin\beta_{1}\sin\alpha_{5}\sin\beta_{5}\sin(\beta_{2}-\alpha_{2})\sin(\alpha_{6}-\beta_{6})(\sin^{2}\alpha_{3}+\sin^{2}\beta_{3})\right]\\
&+N^{2}(\cos^{2}\alpha_{1}+\cos^{2}\beta_{1})(\cos^{2}\alpha_{3}+\cos^{2}\beta_{3})(\cos^{2}\alpha_{5}+\cos^{2}\beta_{5})\\
&=4\rho_{1,16}^{2}\left[(\sin^{2}\alpha_{1}+\sin^{2}\beta_{1})(\sin^{2}\alpha_{3}+\sin^{2}\beta_{3})(\sin^{2}\alpha_{5}+\sin^{2}\beta_{5})\right.\\
&\left.+4\sin\alpha_{1}\sin\beta_{1}\sin\alpha_{3}\sin\beta_{3}\sin(\alpha_{2}-\beta_{2})\sin(\alpha_{4}-\beta_{4})(\sin^{2}\alpha_{5}+\sin^{2}\beta_{5})\right.\\
&\left.+4\sin\alpha_{3}\sin\beta_{3}\sin\alpha_{5}\sin\beta_{5}\sin(\beta_{4}-\alpha_{4})\sin(\alpha_{6}-\beta_{6})(\sin^{2}\alpha_{1}+\sin^{2}\beta_{1})\right.\\
&\left.+4\sin\alpha_{1}\sin\beta_{1}\sin\alpha_{5}\sin\beta_{5}\sin(\beta_{2}-\alpha_{2})\sin(\alpha_{6}-\beta_{6})(\sin^{2}\alpha_{3}+\sin^{2}\beta_{3})\right]\\
&+N^{2}(2-\sin^{2}\alpha_{1}-\sin^{2}\beta_{1})(2-\sin^{2}\alpha_{3}-\sin^{2}\beta_{3})(2-\sin^{2}\alpha_{5}-\sin^{2}\beta_{5}).
\end{aligned}
\end{eqnarray}
\end{widetext}

To calculate the maximum value of $\|\bm{\lambda_{0}}\|^{2}+\|\bm{\lambda_{1}}\|^{2}$, we may suppose that $0\leq\sin\alpha_{1},\sin\beta_{1},\sin\alpha_{3},\allowbreak \sin\beta_{3},\sin\alpha_{5},\sin\beta_{5}\leq1$, $\sin(\alpha_{2}-\beta_{2})=\sin(\alpha_{4}-\beta_{4})=\sin(\beta_{6}-\alpha_{6})=1$. Thus,
\begin{widetext}
\begin{eqnarray}
\begin{aligned}
\mathop{\max}\limits_{\mathbf{a},\mathbf{a'},\mathbf{b},\mathbf{b'},\mathbf{d},\mathbf{d'}}\|\bm{\lambda_{0}}\|^{2}+\|\bm{\lambda_{1}}\|^{2}&=\mathop{\max}\limits_{\alpha_{1},\beta_{1},\alpha_{3},\beta_{3},\alpha_{5},\beta_{5}}
4\rho_{1,16}^{2}\left[(\sin^{2}\alpha_{1}+\sin^{2}\beta_{1})(\sin^{2}\alpha_{3}+\sin^{2}\beta_{3})(\sin^{2}\alpha_{5}+\sin^{2}\beta_{5})\right.\\
&\left.+4\sin\alpha_{1}\sin\beta_{1}\sin\alpha_{3}\sin\beta_{3}(\sin^{2}\alpha_{5}+\sin^{2}\beta_{5})
+4\sin\alpha_{3}\sin\beta_{3}\sin\alpha_{5}\sin\beta_{5}(\sin^{2}\alpha_{1}\right.\\
&\left.+\sin^{2}\beta_{1})
+4\sin\alpha_{1}\sin\beta_{1}\sin\alpha_{5}\sin\beta_{5}(\sin^{2}\alpha_{3}+\sin^{2}\beta_{3})\right]\\
&+N^{2}(2-\sin^{2}\alpha_{1}-\sin^{2}\beta_{1})(2-\sin^{2}\alpha_{3}-\sin^{2}\beta_{3})(2-\sin^{2}\alpha_{5}-\sin^{2}\beta_{5})\\
&=\mathop{\max}\limits_{\alpha_{1},\beta_{1},\alpha_{3},\beta_{3},\alpha_{5},\beta_{5}}4\rho_{1,16}^{2}\delta+N^{2}\delta',
\end{aligned}
\end{eqnarray}
\end{widetext}
where $\delta=(\sin^{2}\alpha_{1}+\sin^{2}\beta_{1})(\sin^{2}\alpha_{3}+\sin^{2}\beta_{3})(\sin^{2}\alpha_{5}+\sin^{2}\beta_{5})
+4\sin\alpha_{1}\sin\beta_{1}\sin\alpha_{3}\sin\beta_{3}(\sin^{2}\alpha_{5}+\sin^{2}\beta_{5})\allowbreak
+4\sin\alpha_{3}\sin\beta_{3}\sin\alpha_{5}\sin\beta_{5}(\sin^{2}\alpha_{1}+\sin^{2}\beta_{1})
+4\sin\alpha_{1}\allowbreak\sin\beta_{1}\sin\alpha_{5}\sin\beta_{5}(\sin^{2}\alpha_{3}+\sin^{2}\beta_{3})$ and
$\delta'=(2-\sin^{2}\alpha_{1}-\sin^{2}\beta_{1})(2-\sin^{2}\alpha_{3}-\sin^{2}\beta_{3})(2-\sin^{2}\alpha_{5}-\sin^{2}\beta_{5})$.
We assert that this maximum value is equal to max$\{128\rho_{1,16}^{2},8N^{2}\}$.

Let $0\leq x,y\leq2$ be two real numbers. Then the inequalities $xy\leq x+y$ and $2xy\leq x^{2}+y^{2}$ evidently holds. Therefore,
\begin{widetext}
\begin{eqnarray}
\begin{aligned}
\delta+4\delta'&=(\sin^{2}\alpha_{1}+\sin^{2}\beta_{1})(\sin^{2}\alpha_{3}+\sin^{2}\beta_{3})(\sin^{2}\alpha_{5}+\sin^{2}\beta_{5})
+4\sin\alpha_{1}\sin\beta_{1}\sin\alpha_{3}\sin\beta_{3}(\sin^{2}\alpha_{5}+\sin^{2}\beta_{5})\\
&+4\sin\alpha_{3}\sin\beta_{3}\sin\alpha_{5}\sin\beta_{5}(\sin^{2}\alpha_{1}+\sin^{2}\beta_{1})
+4\sin\alpha_{1}\sin\beta_{1}\sin\alpha_{5}\sin\beta_{5}(\sin^{2}\alpha_{3}+\sin^{2}\beta_{3})\\
&+4(2-\sin^{2}\alpha_{1}-\sin^{2}\beta_{1})(2-\sin^{2}\alpha_{3}-\sin^{2}\beta_{3})(2-\sin^{2}\alpha_{5}-\sin^{2}\beta_{5})\\
&=32-3(\sin^{2}\alpha_{1}+\sin^{2}\beta_{1})(\sin^{2}\alpha_{3}+\sin^{2}\beta_{3})(\sin^{2}\alpha_{5}+\sin^{2}\beta_{5})+8\left[(\sin^{2}\alpha_{1}+\sin^{2}\beta_{1})(\sin^{2}\alpha_{3}+\sin^{2}\beta_{3})\right.\\
&\left.+(\sin^{2}\alpha_{3}+\sin^{2}\beta_{3})(\sin^{2}\alpha_{5}+\sin^{2}\beta_{5})+(\sin^{2}\alpha_{1}+\sin^{2}\beta_{1})(\sin^{2}\alpha_{5}+\sin^{2}\beta_{5})\right]\\
&-16(\sin^{2}\alpha_{1}+\sin^{2}\beta_{1}+\sin^{2}\alpha_{3}+\sin^{2}\beta_{3}+\sin^{2}\alpha_{5}+\sin^{2}\beta_{5})+4\sin\alpha_{1}\sin\beta_{1}\sin\alpha_{3}\sin\beta_{3}(\sin^{2}\alpha_{5}+\sin^{2}\beta_{5})\\
&+4\sin\alpha_{3}\sin\beta_{3}\sin\alpha_{5}\sin\beta_{5}(\sin^{2}\alpha_{1}+\sin^{2}\beta_{1})
+4\sin\alpha_{1}\sin\beta_{1}\sin\alpha_{5}\sin\beta_{5}(\sin^{2}\alpha_{3}+\sin^{2}\beta_{3})\\
&\leq32.
\end{aligned}
\end{eqnarray}
\end{widetext}
If $128\rho_{1,16}^{2}\geq 8N^{2}$, then
\begin{eqnarray}
\begin{aligned}
128\rho_{1,16}^{2}-(4\rho_{1,16}^{2}\delta+N^{2}\delta')&=4\rho_{1,16}^{2}(32-\delta)-N^{2}\delta'\\
&\geq16\rho_{1,16}^{2}\delta'-N^{2}\delta'\\
&\geq0.
\end{aligned}
\end{eqnarray}
If $8N^{2}\geq128\rho_{1,16}^{2}$, then
\begin{eqnarray}
\begin{aligned}
8N^{2}-(4\rho_{1,16}^{2}\delta+N^{2}\delta')&=N^{2}(8-\delta')-4\rho_{1,16}^{2}\delta\\
&\geq\frac{1}{4}N^{2}\delta-4\rho_{1,16}^{2}\delta\\
&\geq0.
\end{aligned}
\end{eqnarray}
Besides,
\begin{widetext}
\begin{eqnarray}
\begin{aligned}
\langle\bm{\lambda_{0}},\bm{\lambda_{1}}\rangle&=(\mathbf{d'}T_{a'b'}-\mathbf{d'}T_{ab}-\mathbf{d}T_{a'b}-\mathbf{d}T_{ab'})(T_{ab}\mathbf{d}-T_{a'b'}\mathbf{d}-T_{a'b}\mathbf{d'}-T_{ab'}\mathbf{d'})\\
&=4\rho_{1,16}^{2}\left[(a_{1}^{\prime2}+a_{2}^{\prime2}-a_{1}^{2}-a_{2}^{2})(b_{1}^{2}+b_{2}^{2}-b_{1}^{\prime2}-b_{2}^{\prime2})(d_{1}d'_{1}+d_{2}d'_{2})\right.\\
&\left.+4(a_{1}a'_{1}+a_{2}a'_{2})(b_{1}b'_{1}+b_{2}b'_{2})(d_{1}+d'_{1}+d_{2}d'_{2})+(a_{1}^{2}+a_{2}^{2}-a_{1}^{\prime2}-a_{2}^{\prime2})(d_{1}^{2}+d_{2}^{2}-d_{1}^{\prime2}-d_{2}^{\prime2})(b_{1}b'_{1}+b_{2}b'_{2})\right.\\
&\left.+(b_{1}^{2}+b_{2}^{2}-b_{1}^{\prime2}-b_{2}^{\prime2})(d_{1}^{2}+d_{2}^{2}-d_{1}^{\prime2}-d_{2}^{\prime2})(a_{1}a'_{1}+a_{2}a'_{2})\right]+N^{2}\left[b_{3}b'_{3}(a_{3}^{2}-a_{3}^{\prime2})+(d_{3}^{2}-d_{3}^{\prime2})\right.\\
&\left.+a_{3}a'_{3}(b_{3}^{2}-b_{3}^{\prime2})+(d_{3}^{2}-d_{3}^{\prime2})+4a_{3}a'_{3}b_{3}b'_{3}d_{3}d'_{3}\right]\\
&=4\rho_{1,16}^{2}\left[\sin\alpha_{5}\sin\beta_{5}\cos(\alpha_{6}-\beta_{6})(\sin^{2}\beta_{1}-\sin^{2}\alpha_{1})(\sin^{2}\alpha_{3}-\sin^{2}\beta_{3})\right.\\
&\left.+4\sin\alpha_{1}\sin\beta_{1}\sin\alpha_{3}\sin\beta_{3}\sin\alpha_{5}\sin\beta_{5}\cos(\alpha_{2}-\beta_{2})\cos(\alpha_{4}-\beta_{4})\cos(\alpha_{6}-\beta_{6})\right.\\
&\left.+\sin\alpha_{1}\sin\beta_{1}\cos(\alpha_{2}-\beta_{2})(\sin^{2}\alpha_{5}-\sin^{2}\beta_{5})(\sin^{2}\alpha_{3}-\sin^{2}\beta_{3})\right.\\
&\left.+\sin\alpha_{3}\sin\beta_{3}\cos(\alpha_{4}-\beta_{4})(\sin^{2}\alpha_{5}-\sin^{2}\beta_{5})(\sin^{2}\alpha_{1}-\sin^{2}\beta_{1})\right]\\
&+N^{2}\left[\cos\alpha_{3}\cos\beta_{3}(\cos^{2}\alpha_{5}-\cos^{2}\beta_{5})(\cos^{2}\alpha_{1}-\cos^{2}\beta_{1})+\cos\alpha_{1}\cos\beta_{1}(\cos^{2}\alpha_{5}-\cos^{2}\beta_{5})(\cos^{2}\alpha_{3}-\cos^{2}\beta_{3})\right.\\
&\left.+\cos\alpha_{5}\cos\beta_{5}(\cos^{2}\alpha_{1}-\cos^{2}\beta_{1})(\cos^{2}\beta_{3}-\cos^{2}\alpha_{3})+4\cos\alpha_{1}\cos\beta_{1}\cos\alpha_{3}\cos\beta_{3}\cos\alpha_{5}\cos\beta_{5}\right].
\end{aligned}
\end{eqnarray}
\end{widetext}

As it is assumed that $\sin(\alpha_{2}-\beta_{2})=\sin(\alpha_{4}-\beta_{4})=\sin(\beta_{6}-\alpha_{6})=1$, $\cos(\alpha_{2}-\beta_{2})=\cos(\alpha_{4}-\beta_{4})=\cos(\alpha_{6}-\beta_{6})=0$. Thus, $\langle\bm{\lambda_{0}},\bm{\lambda_{1}}\rangle=0$ when $\sin\alpha_{1}=\sin\beta_{1}=1$.
The maximum value of $\|\bm{\lambda_{0}}\|^{2}+\|\bm{\lambda_{1}}\|^{2}$ is attained for $\bm{\lambda_{0}}$ and $\bm{\lambda_{1}}$ with $\bm{\lambda_{0}}\bot\bm{\lambda_{1}}$. According to Proposition \ref{Pr2}, we have
\begin{eqnarray}
\begin{aligned}
S(\rho)=max\left\{16\sqrt{2}|\rho_{1,16}|,4\sqrt{2}|N|\right\},
\end{aligned}
\end{eqnarray}
where
$N=\rho_{1,1}-\rho_{2,2}-\rho_{3,3}+\rho_{4,4}-\rho_{5,5}+\rho_{6,6}+\rho_{7,7}-\rho_{8,8}-\rho_{9,9}+\rho_{10,10}+\rho_{11,11}-\rho_{12,12}+\rho_{13,13}-\rho_{14,14}-\rho_{15,15}+\rho_{16,16}$.

\section{APPENDIX B: sub-matrices $M$, $N$ and $V$}
\label{App2}
From \cite{Phy2024}, the specific expression of the sub-matrix $M$ is given by
\begin{eqnarray*}
M=\alpha^{2}
\left (
\begin{array}{cccccc}
\frac{1}{(e^{-\frac{\omega}{T}}+1)^{n}}       &       &           &        &       &           \\
             &\ddots                                 &          &               &       &           \\
             &       & \frac{1}{(e^{\frac{\omega}{T}}+1)^{n}}          &              &       &           \\
             &       &                &0       &           & \\
             &       &                &        &\ddots     & \\
             &       &                &        &           &0 \\
\end{array}
\right )
\end{eqnarray*}
in the $2^{n}$ basis $\{|\overline{0}0\ldots00\rangle,\ldots,|\overline{0}1\ldots11\rangle\}$, where the base corresponding to the element $\frac{\alpha^{2}}{(e^{-\frac{\omega}{T}}+1)^{n-i}(e^{\frac{\omega}{T}}+1)^{i}}$ contains $i$ $``1"$.
Note that the element at the position $(2^{n},2^{n})$ is $\frac{\alpha^{2}}{(e^{\frac{\omega}{T}}+1)^{n}}$ when $n=3$. In this case the diagonal elements of $M$ are non-zero.

The sub-matrix $N$ can be written as
\begin{eqnarray*}
N=
\left (
\begin{array}{ccccccc}
0       &           &        &              &       &       & \\
        &\ddots     &        &              &       &       & \\
        &           &0       &              &       &       & \\
        &           &        &1-\alpha^{2}  &       &       & \\
        &           &        &              &0      &       & \\
        &           &        &              &       &\ddots & \\
        &           &        &              &       &       &0\\
\end{array}
\right )
\end{eqnarray*}
and the $V$ is give by
\begin{eqnarray*}
V=
\left (
\begin{array}{ccccccc}
        &           &        &              &       &       &0\\
        &           &        &              &       & \iddots  & \\
        &           &        &              &0      &       & \\
        &           &        & \frac{\alpha\sqrt{1-\alpha^{2}}}{\sqrt{(e^{-\frac{\omega}{T}}+1)^{p}(e^{\frac{\omega}{T}}+1)^{q}}}  &       &       & \\
        &           &0       &              &       &       & \\
        & \iddots     &        &              &       &       & \\
0       &           &        &              &       &       & \\
\end{array}
\right ).
\end{eqnarray*}

\section{APPENDIX C: sub-matrixes $M_{A}$, $M_{B}$ and $M_{X}$}
\label{App3}
The $8\times8$ sub-matrix $M_{A}$ can be written as~\cite{Wu2024},
\begin{eqnarray*}
M_{A}=\alpha^{2}
\left (
\begin{array}{cccc}
\rho_{1,1}    &             &           & \\
             &\rho_{2,2}    &           & \\
             &             &\ddots     & \\
             &             &           &\rho_{8,8}\\
\end{array}
\right ),
\end{eqnarray*}
where $\rho_{1,1}=\cos^{2n}r\cos^{2m}\omega$, $\rho_{2,2}=\cos^{2n}r\cos^{2(m-1)}\omega \cdot\sin^{2}\omega$ and $\rho_{8,8}=\cos^{2}r\sin^{2(n-1)}r\sin^{2m}\omega$.

The sub-matrices $M_{X}$ and $M_{B}$ are given by
\begin{eqnarray*}
M_{X}=\alpha\sqrt{1-\alpha^{2}}
\left (
\begin{array}{cccc}
             &                                               &           &\cos^{n}r\cos^{m}\omega \\
             &                                               &0          & \\
             & \iddots             &           & \\
0            &                                               &           & \\
\end{array}
\right )
\end{eqnarray*}
and
\begin{eqnarray*}
M_{B}=
\left (
\begin{array}{ccccc}
\rho_{9,9}    &               &           &              &\\
             &\rho_{10,10}    &           &              &\\
             &                &\ddots     &              &\\
             &                &           &\rho_{15,15}  &\\
             &                &           &              &\rho_{16,16}\\
\end{array}
\right ),
\end{eqnarray*}
respectively, where $\rho_{9,9}=\alpha^{2}\cos^{2(n-1)}r\sin^{2}r\cos^{2m}\omega$, $\rho_{15,15}=\alpha^{2}\sin^{2n}r\cos^{2}\omega\sin^{2(m-1)}\omega$,
$\rho_{16,16}=\alpha^{2}\sin^{2n}r\sin^{2m}\omega+1-\alpha^{2}$ and $\rho_{10,10}=\alpha^{2}\cos^{2(n-1)}r\sin^{2}r\cos^{2(m-1)}\omega\sin^{2}\omega$.

\end{document}